\documentclass[showpacs,twoside,twocolumn,english,aps,prl]{revtex4}
\usepackage[T1]{fontenc}
\usepackage[latin1]{inputenc}
\usepackage{graphicx}
\usepackage{amssymb}
\usepackage{verbatim}

\makeatletter

\newcommand{\ba}{\begin{eqnarray}}
\newcommand{\ea}{\end{eqnarray}}

\usepackage{epic,eepic}
\usepackage{babel}
\makeatother

\begin{document}

\title{Two-nucleon transfer reactions uphold 
supersymmetry in atomic nuclei}

\author{J. Barea}
\email{barea@nucleares.unam.mx}
\affiliation{Instituto de Ciencias Nucleares, 
Universidad Nacional Aut\'onoma de M\'exico, 
AP 70-543, 04510 M\'exico, D.F. M\'exico}

\author{R. Bijker}
\email{bijker@nucleares.unam.mx}
\affiliation{Instituto de Ciencias Nucleares, 
Universidad Nacional Aut\'onoma de M\'exico, 
AP 70-543, 04510 M\'exico, D.F. M\'exico}

\author{A. Frank}
\email{frank@nucleares.unam.mx}
\affiliation{Instituto de Ciencias Nucleares, 
Universidad Nacional Aut\'onoma de M\'exico, 
AP 70-543, 04510 M\'exico, D.F. M\'exico}

\date{April 15, 2005}

\begin{abstract}
The spectroscopic strengths of two-nucleon transfer reactions constitute a 
stringent test for two-nucleon correlations in the nuclear wave functions. 
A set of closed analytic expressions for ratios of spectroscopic factors 
is derived in the framework of nuclear supersymmetry. These ratios are 
parameter independent and provide a direct test of the wave functions. 
A comparison between the recently measured 
$^{198}\textrm{Hg}(\vec{d},\alpha){}^{196}\textrm{Au}$ reaction 
and the predictions from the nuclear quartet supersymmetry lends further 
support to the validity of supersymmetry in nuclear physics. 
\end{abstract}


\maketitle

The Interacting Boson Model (IBM) and its extensions have provided 
a bridge between single-particle and collective behavior in nuclei, 
based on the approximate bosonic nature of pairs of 
identical nucleons that dominate the dynamics of valence nucleons and 
that arise from the underlying nuclear forces. This is similar to 
the BCS theory of semi-conductors with its coupling of electrons
to spin-zero Cooper pairs which leads to collective behavior and 
to superconductivity. The conceptual basis of the IBM has led to a 
unified description of the collective properties of medium and heavy 
mass even-even nuclei, pictured in this framework as belonging to 
transitional regions between various dynamical symmetries 
\cite{IBM}. 

Odd-mass nuclei were also analyzed from this point of view, by 
incorporating the degrees of freedom of a single fermion 
\cite{IBFM}. In 1980, Iachello suggested a 
simultaneous description of even-even and odd-mass nuclei 
through the introduction of a superalgebra, with energy levels in both
nuclei belonging to the same (super)multiplet \cite{FI}. In essence,
this proposal is based on the fact that even-even nuclei behave as
(composite) bosons while odd-mass ones behave as approximate
fermions. At the appropriate length scales their states can be
viewed as elementary. The bold and far-reaching idea was then put
forward that both these nuclei can be embedded into a single
conceptual framework, relating boson-boson and boson-fermion
interactions in a precise way. 

The concept of nuclear supersymmetry was extended in 1985 to include the  
neutron-proton degree of freedom \cite{quartet}. In the new framework, 
a supermultiplet consists of an even-even, an odd-proton, an odd-neutron 
and an odd-odd nucleus. Spectroscopic studies of heavy odd-odd nuclei 
are very difficult due the high density of states. Almost 15 years 
after the prediction of the spectrum of the odd-odd nucleus by nuclear 
supersymmetry, it was shown experimentally that the observed spectrum 
of the nucleus $^{196}$Au is amazingly close to the theoretical one 
\cite{metz}. 

However, transfer reactions provide a far more sensitive test of the wave 
functions than do energies. In particular, two-nucleon transfer reactions 
constitute a powerful tool in nuclear structure research \cite{Glendenning}. 
In contrast to the better studied one-nucleon transfer reactions where the 
single-particle content of the states of the final nucleus is scrutinized, 
two-nucleon transfer reactions probe the structure of these states in a 
more subtle way through the exploration of two-nucleon correlations that 
may be present. The spectroscopic strengths of the two-nucleon transfer 
reaction depend on two factors: the similarity between the states 
in the initial and final nucleus which differ by two nucleons,
and the correlation of the tranferred pair of nucleons. 
The information extracted through these reactions supply a challenging 
test of the calculated wave functions for any nuclear structure model.

The purpose of this Letter is to study two-nucleon transfer reactions in 
the $U_{\nu}(6/12)\otimes U_{\pi}(6/4)$ supersymmetry via selection rules 
and spectroscopic strengths and to test the predictions against the recent 
data on the $^{198}\textrm{Hg}(\vec{d},\alpha){}^{196}\textrm{Au}$ reaction 
obtained in 2004 \cite{wirth}. This reaction involves the transfer of a 
proton-neutron pair, and hence measures the neutron-proton correlation 
in the odd-odd nucleus. 

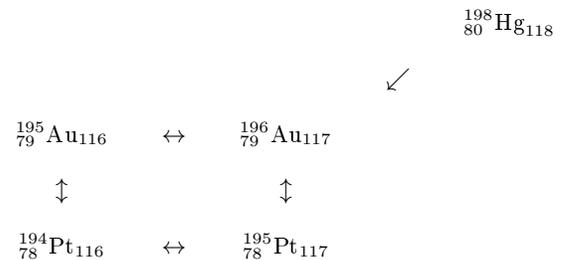
\begin{table}[h]
\centering
\ba
\begin{array}{ccccc}
& & & & ^{198}_{ 80}\mbox{Hg}_{118} \\
& & & & \\
& & & \hspace{0.5cm} \swarrow \hspace{0.5cm} & \\
& & & & \\
^{195}_{ 79}\mbox{Au}_{116} & \hspace{0.5cm} \leftrightarrow \hspace{0.5cm} & 
^{196}_{ 79}\mbox{Au}_{117} & & \\
& & & & \\
\updownarrow & & \updownarrow & & \\
& & & & \\
^{194}_{ 78}\mbox{Pt}_{116} & \hspace{0.5cm} \leftrightarrow \hspace{0.5cm} & 
^{195}_{ 78}\mbox{Pt}_{117} & & \\
%& & & & \\
\end{array}
\label{magic}
\nonumber
\ea
\end{table}

The final odd-odd nucleus $^{196}$Au has been suggested as a member 
of a supersymmetric quartet of nuclei \cite{quartet} of the 
$U_{\nu}(6/12)\otimes U_{\pi}(6/4)$ dynamical supersymmetry 
in which the  
odd neutron is allowed to occupy the $2\nu f_{5/2}$, $3\nu p_{3/2}$ and 
$3\nu p_{1/2}$ orbits of the 82-126 shell and the odd proton the 
$2\pi d_{3/2}$ level of the 50-82 shell. It incorporates both the $U(6/4)$ 
scheme \cite{FI} for the even-even and odd-proton nuclei and the $U(6/12)$ 
scheme \cite{baha} for the even-even and odd-neutron nuclei in the Pt-Au 
mass region. 
%As a consequence the energy spectrum of the odd-odd 
%nucleus $^{196}$Au was predicted \cite{quartet} and confirmed later  
%experimentally \cite{metz}. 
In this extended scheme which 
includes the neutron-proton degree of freedom, the four nuclei 
$^{194,195}$Pt and $^{195,196}$Au form a supersymmetric quartet.  

The relevant subgroup chain of $U(6/12)_{\nu} \otimes U(6/4)_{\pi}$ for 
the Pt and Au nuclei is given by \cite{quartet} 
\ba
U(6/12)_{\nu} &\otimes& U(6/4)_{\pi} 
\nonumber\\
&\supset& U^{B_{\nu}}(6) \otimes U^{F_{\nu}}(12) \otimes 
U^{B_{\pi}}(6) \otimes U^{F_{\pi}}(4) 
\nonumber\\
&\supset& U^B(6) \otimes U^{F_{\nu}}(6) \otimes U^{F_{\nu}}(2) \otimes 
U^{F_{\pi}}(4) 
\nonumber\\
&\supset& U^{BF_{\nu}}(6) \otimes U^{F_{\nu}}(2) \otimes U^{F_{\pi}}(4)
\nonumber\\
&\supset& SO^{BF_{\nu}}(6) \otimes U^{F_{\nu}}(2) \otimes SU^{F_{\pi}}(4)
\nonumber\\
&\supset& Spin(6) \otimes U^{F_{\nu}}(2) 
\nonumber\\
&\supset& Spin(5) \otimes U^{F_{\nu}}(2) 
\nonumber\\
&\supset& Spin(3) \otimes SU^{F_{\nu}}(2) 
\nonumber\\
&\supset& SU(2) ~.
\label{chain}
\ea
In a dynamical supersymmetry the Hamiltonian is expressed 
in terms of Casimir invariants of the groups appearing in the chain 
of Eq.~(\ref{chain}) leading to a closed form for the energy spectrum 
and a direct correlation between the wave functions of the four nuclei 
that make up the quartet.  

For simplicity, the ground state wave function of the initial nucleus 
$^{198}$Hg is taken to be that of the $SO(6)$ limit of the IBM \cite{IBM} 
\ba
\left| ^{198}\mbox{Hg} \right> &=& 
\left|[N_{\nu}],[N_{\pi}];[N],(N,0,0),(0,0),0 \right> ~,
\label{wfhg}
\ea
where $N=N_{\nu}+N_{\pi}$ is the total number of bosons. 
Its parity is positive. 
In the nuclear supersymmetry classification scheme,  the wave functions 
of the final nucleus $^{196}$Au have a more complicated 
structure since they consist of a bosonic part characterized by the same 
number of proton and neutron bosons as the initial nucleus $^{198}$Hg, 
and a fermionic part for the proton orbit $j_{\pi}=3/2$ and the neutron 
orbits $j_{\nu}=1/2$, $3/2$, $5/2$ characterized by the labels 
\ba
\pi &:& \left| (\frac{1}{2},\frac{1}{2},\frac{1}{2}),
(\frac{1}{2},\frac{1}{2}),j_{\pi}=\frac{3}{2} \right> ~, 
\nonumber\\
\nu &:& \left| [1],(1,0,0),(\tau,0),2\tau,\frac{1}{2};j_{\nu} \right> ~,
\label{orbits}
\ea
with $\tau=0,1$. The labels of the proton orbital are those of 
the spinor representations of $SO(6)$ and $SO(5)$ \cite{FI}.  
The neutron orbitals are decomposed into a pseudo-orbital part 
$k=2\tau$ (with $\tau=0,1$) and a spin part $s=1/2$. The pseudo-orbital 
angular momenta span the six-dimensional representations  
$[1]$ and $(1,0,0)$ of $U(6)$ and $SO(6)$, respectively, which contain 
$(\tau,0)=(0,0)$ and $(1,0)$ of $SO(5)$ \cite{baha}. 
The wave functions of $^{196}$Au are obtained by combining those 
of the even-even nucleus of Eq.~(\ref{wfhg}) with the 
single-particle wave functions of Eq.~(\ref{orbits}) into 
\ba
\left| ^{196}\mbox{Au} \right> \;=\; \left|[N_{\nu}],[N_{\pi}];[N],
[1]_{\nu};[N_1,N_2],(\Sigma_1,\Sigma_2,0), \right.
\nonumber\\
\left. \left(\frac{1}{2},\frac{1}{2},\frac{1}{2}\right)_{\pi};
(\sigma_1,\sigma_2,\sigma_3),(\tau_1,\tau_2),J',\frac{1}{2};J \right> ~, 
\label{wfau}
\ea
where the labels denote the irreducible representations of the groups 
appearing in Eq.~(\ref{chain}) \cite{quartet}. 
Due to the choice of the single-particle orbits, the parity of the 
states in Eq.~(\ref{wfau}) is odd. 

In first order, the form of the two-nucleon transfer operator for the 
$(\vec{d},\alpha)$ reaction is simply given by 
\ba
( a_{j_{\nu}}^{\dagger} a_{j_{\pi}}^{\dagger} )^{(\lambda)} ~. 
\label{operator}
\ea
In the IBM and its extensions the number of particles 
corresponds to the number of valence nucleons in the first half 
of the major shell and to holes in the second half. Therefore, in the 
present appliciation the creation operators in Eq.~(\ref{operator}) 
correspond to holes. 
The selection rules can be determined from the tensorial character of the 
proton and neutron orbits of Eq.~(\ref{orbits}). As a result, the transfer 
operator of Eq.~(\ref{operator}) can be expanded in terms of three tensor 
operators $T^{(s_1,s_2,s_3)}_{(t_1,t_2),J',1/2;J}$ where $(s_1,s_2,s_3)$ 
denotes the tensorial character under $Spin(6)$, $(t_1,t_2)$ under 
$Spin(5)$, $J'$ under $Spin(2)$ and $J$ under $SU(2)$
\begin{eqnarray}
T_{1} &=& T_{(\frac{1}{2},\frac{1}{2}),J',\frac{1}{2};J}
^{(\frac{1}{2},\frac{1}{2},-\frac{1}{2})} ~, 
\nonumber\\
T_{2} &=& T_{(\frac{1}{2},\frac{1}{2}),J',\frac{1}{2};J}
^{(\frac{3}{2},\frac{1}{2},\frac{1}{2})} ~, 
\nonumber\\
T_{3} &=& T_{(\frac{3}{2},\frac{1}{2}),J',\frac{1}{2};J}
^{(\frac{3}{2},\frac{1}{2},\frac{1}{2})} ~. 
\label{tensor}
\end{eqnarray}
The allowed values of $(\sigma_1,\sigma_2,\sigma_3)$ and $(\tau_1,\tau_2)$ 
of the wave functions of $^{196}$Au are presented in Table~\ref{rules}. 

\begin{table}
\centering
\caption[]{States in $^{196}$Au that can be excited from the ground state 
in $^{198}$Hg by the tensor operators $T_1$, $T_2$ and $T_3$ of 
Eq.~(\ref{tensor}) for two-nucleon transfer reactions.} 
\label{rules}
\vspace{15pt}
\begin{tabular}{ccccc}
\hline
& & & & \\
$(\sigma_1,\sigma_2,\sigma_3)$ & $(\tau_1,\tau_2)$ & $T_1$ & $T_2$ & $T_3$ \\
& & & & \\
\hline
& & & & \\
$(N\pm\frac{3}{2},\frac{1}{2},\pm\frac{1}{2})$ 
& $(\frac{1}{2},\frac{1}{2})$ & & $\surd$ & \\ 
& $(\frac{3}{2},\frac{1}{2})$ & & & $\surd$ \\
$(N\pm\frac{1}{2},\frac{1}{2},\mp\frac{1}{2})$ 
& $(\frac{1}{2},\frac{1}{2})$ & $\surd$ & $\surd$ & \\ 
& $(\frac{3}{2},\frac{1}{2})$ & & & $\surd$ \\
$(N\pm\frac{1}{2},\frac{3}{2},\pm\frac{1}{2})$ 
& $(\frac{3}{2},\frac{1}{2})$ & & & $\surd$ \\
& & & & \\
\hline
\end{tabular}
\label{tau}
\end{table}

The matrix elements of the transfer operator of Eq.~(\ref{operator}) 
can be derived in closed analytic form by using standard tensor algebra 
\ba
\left< ^{196}\mbox{Au} \right\| 
( a_{j_{\nu}}^{\dagger} a_{j_{\pi}}^{\dagger} )^{(\lambda)}
\left\| ^{198}\mbox{Hg} \right> \hspace{2cm}
\nonumber\\
\;=\; \delta_{\lambda,J} (-)^{j_{\nu}+J'} \hat{J} 
\hat{j_{\nu}} \hat{J'} \left\{ \begin{array}{ccc} 
2\tau & \frac{1}{2} & j_{\nu} \\ 
& & \\ J & \frac{3}{2} & J' \end{array} \right\} 
\nonumber\\
\left< \begin{array}{cc} [N] & [1] \\ & \\ (N,0,0) & (1,0,0) \end{array} 
\right| \left. \begin{array}{c} [N_1,N_2] \\ \\ (\Sigma_1,\Sigma_2,0) 
\end{array} \right> 
\nonumber\\
\left< \begin{array}{cc} (\Sigma_1,\Sigma_2,0) 
& (\frac{1}{2},\frac{1}{2},\frac{1}{2}) \\ & \\
(\tau,0),2\tau & (\frac{1}{2},\frac{1}{2}),\frac{3}{2} \end{array} 
\right| \left. \begin{array}{c} (\sigma_1,\sigma_2,\sigma_3) \\ \\ 
(\tau_1,\tau_2),J' \end{array} \right> 
\nonumber\\
\left< \begin{array}{cc} (N,0,0) & (1,0,0) \\ & \\
(0,0),0 & (\tau,0),2\tau \end{array} 
\right| \left. \begin{array}{c} (\Sigma_1,\Sigma_2,0) \\ \\ (\tau,0),2\tau 
\end{array} \right> ~,
\label{rme}
\ea
with $\hat{L}=\sqrt{2L+1}$. The coefficients in brackets $<|>$ 
denote isoscalar factors \cite{IK,IFS,BI,PvI,BBF}. The explicit results 
will be published in a separate article \cite{BBF}. 

The $^{198}\textrm{Hg}(\vec{d},\alpha){}^{196}\textrm{Au}$ reaction is
characterized by the transfer of a correlated neutron-proton pair with 
spin $S=1$. Since the angular momentum of the ground state of 
$^{198}\textrm{Hg}$ is zero, the transfered total angular momentum 
$\lambda$ is equal to the angular momentum $J$ of the final 
state of $^{196}\textrm{Au}$. Thus for each value of $\lambda=J$ there 
are three different transfers corresponding to $L=J-1$, $J$ and $J+1$. 
Since the initial and final states have opposite parity, parity conservation 
limits the allowed values of $L$ to be odd. The transferred angular momentum 
and parity of the two-nucleon transfer operator in Eq.~(\ref{operator}) is 
$J^{\pi}=0^{-}$, $1^{-}$, $2^{-}$, $3^{-}$, $4^{-}$. 
The $L$ transfer with total angular momentum $J$ is denoted as $L_{J}$. 
The seven possibilities are $P_{0}$, $P_{1}$, $P_{2}$, 
$F_{2}$, $F_{3}$, $F_{4}$ and $H_{4}$. 

The experimental values of the spectroscopic strengths $G_{LJ}$ for the 
transfer of a neutron-proton pair were determined from the measurement 
of the angular distributions of the differential cross section and the 
analyzing power of the $^{198}$Hg$(\vec{d},\alpha){}^{196}$Au reaction 
\cite{wirth}. Theoretically, the spectroscopic strengths can be written 
as
\begin{equation}
G_{LJ} = \left| \sum_{j_{\nu} j_{\pi}} g_{j_{\nu}j_{\pi}}^{LJ} 
\left< ^{196}\mbox{Au} \right\| 
( a_{j_{\nu}}^{\dagger} a_{j_{\pi}}^{\dagger} )^{(\lambda)}
\left\| ^{198}\mbox{Hg} \right> \right|^2 ~, 
\end{equation}
where the coefficients $g_{j_{\nu}j_{\pi}}^{LJ}$ contain factors that 
arise from 
the reaction mechanism for two-nucleon transfer reactions, such as a 
$9-j$ symbol for a change of angular momentum coupling from $jj$ to $LS$ 
coupling and a Talmi-Moshinksy bracket for the transformation 
to relative and center-of-mass coordinates of the transferred 
nucleons \cite{Glendenning}. The nuclear structure part is contained in 
the reduced matrix elements of Eq.~(\ref{rme}). 

In order to compare with experimental data we calculate 
for each combination of $L_J$ the relative strengths from the ratio 
\begin{equation}
R_{LJ}=G_{LJ}/G_{LJ}^{\textrm{ref}} ~,
\end{equation} 
where $G_{LJ}^{\textrm{ref}}$ is the spectroscopic strength of the
reference state for a particular $L_{J}$ transfer. 

The tensorial character of the transfer operator of Eq.~(\ref{operator}) 
shows that in the supersymmetry scheme only states in $^{196}$Au with 
$(\tau_1,\tau_2)=(\frac{3}{2},\frac{1}{2})$ and 
$(\frac{1}{2},\frac{1}{2})$ can be excited (see Table~\ref{tau}). 
The angular momentum states belonging to 
$(\frac{3}{2},\frac{1}{2})$ have $J'=\frac{1}{2}$, $\frac{5}{2}$, 
$\frac{7}{2}$ and $J=J'\pm \frac{1}{2}$. Table~\ref{tau} shows that they 
can only be excited by the tensor operator $T_3$. 
Therefore, the ratios of spectroscopic strengths to these states 
provide a direct test of the nuclear wave functions, since they do not 
depend on the coefficients $g_{j_{\nu}j_{\pi}}$, but only on the nuclear 
structure part, {\it i.e.} the reduced matrix elements of $T_3$. 
If we take the states with 
$[N_1,N_2]=[5,1]$, $(\Sigma_1,\Sigma_2,0)=(5,1,0)$ and 
$(\sigma_1,\sigma_2,\sigma_3)=(\frac{11}{2},\frac{3}{2},\frac{1}{2})$ 
as reference states, we find 
\begin{equation} 
R_{LJ}=\frac{N+4}{15N} ~,
\end{equation}
for $[5,1]$, $(5,1,0)$, $(\frac{11}{2},\frac{1}{2},-\frac{1}{2})$ and
\begin{equation}
R_{LJ}=\frac{2(N+4)(N+6)}{15N(N+3)} ~,
\end{equation} 
for $[6,0]$, $(6,0,0)$, $(\frac{13}{2},\frac{1}{2},\frac{1}{2})$.  
The numerical values are 0.12 and 0.33, respectively (for $N=5$). 
The angular momentum states belonging to $(\frac{1}{2},\frac{1}{2})$ have 
$J'=\frac{3}{2}$ and $J=J'\pm \frac{1}{2}$. Table~\ref{tau} shows that they 
can be excited by the tensor operators $T_1$ and $T_2$. For these 
states the ratios $R_{LJ}$ depend both on the reaction and structure 
part. 

\begin{figure*}
\begin{center}
\includegraphics[scale=0.45]{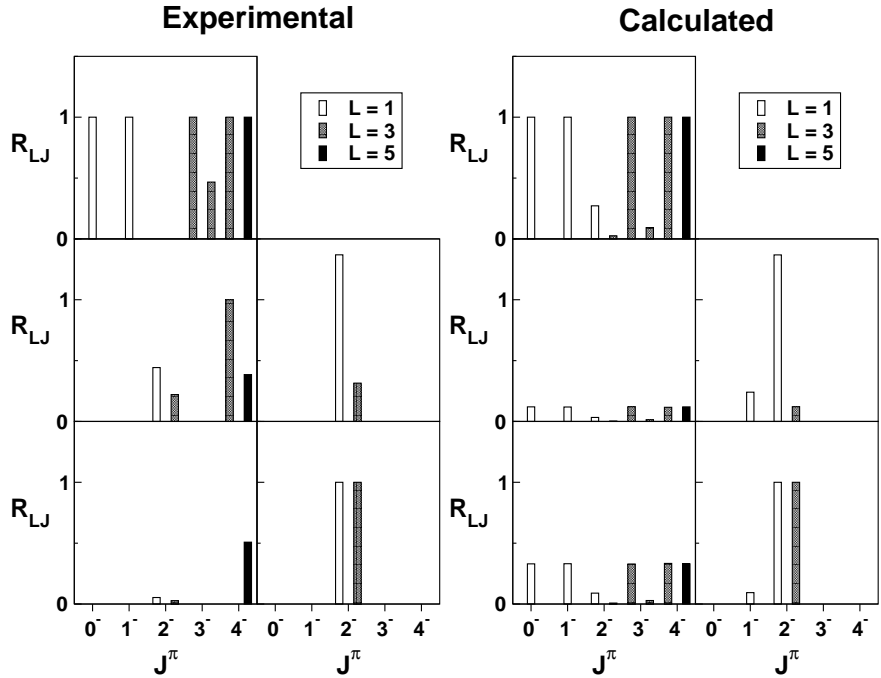}
\end{center}
\caption{Ratios of spectroscopic strengths. The wave functions 
of the final nucleus $^{196}$Au are given by Eq.~(\protect\ref{wfau}). 
The two columns in each frame correspond to states with $Spin(5)$ labels 
$(\tau_1,\tau_2)=(\frac{3}{2},\frac{1}{2})$ and 
$(\frac{1}{2},\frac{1}{2})$, respectively. The rows are 
characterized by the labels $[N_1,N_2]$, $(\Sigma_1,\Sigma_2,0)$, 
$(\sigma_1,\sigma_2,\sigma_3)$. From bottom to top we have (i)  
$[6,0]$, $(6,0,0)$, $(\frac{13}{2},\frac{1}{2}, \frac{1}{2})$, (ii) 
$[5,1]$, $(5,1,0)$, $(\frac{11}{2},\frac{1}{2},-\frac{1}{2})$ and (iii) 
$[5,1]$, $(5,1,0)$, $(\frac{11}{2},\frac{3}{2}, \frac{1}{2})$.}
\label{ratios}
\end{figure*}

In Fig.~\ref{ratios} we show the experimental and calculated
ratios $R_{LJ}$. The reference states can easily be identified 
since they are normalized to one. 
The $P_{0}$, $P_{1}$, $F_{3}$, $F_{4}$ and $H_{4}$ transfers 
are normalized to the states assigned as 
$[5,1]$, $(5,1,0)$, $(\frac{11}{2},\frac{3}{2},\frac{1}{2})$, 
$(\frac{3}{2},\frac{1}{2})$, whereas 
the $P_{2}$ and $F_{2}$ transfers to the 
$[6,0]$, $(6,0,0)$, $(\frac{13}{2},\frac{1}{2},\frac{1}{2})$, 
$(\frac{1}{2},\frac{1}{2})$ states.

We observe in general that there is good overall agreement between 
the experimental and theoretical values, especially if we take into
account the simple form of the operator in the calculation of the 
two-nucleon transfer reaction intensities. We can see that large ratios 
are well reproduced except for one related to a $4^{-}$ state and that 
all small ratios are consistent with the experimental data.

These results have led us to a change in the assignment used
previously for the $2^-$ state at 166.6(5) keV \cite{au196}. 
It is now associated to the theoretical state with labels
$[5,1]$, $(5,1,0)$, $(\frac{11}{2},\frac{1}{2},-\frac{1}{2})$, 
$(\frac{1}{2},\frac{1}{2})$.  

In conclusion, we have studied the two-nucleon pickup reaction
$^{198}\textrm{Hg}(\vec{d},\alpha){}^{196}\textrm{Au}$ as a test 
of the nuclear supersymmetry scheme proposed for the Pt-Au region 
\cite{quartet,metz}. Two-nucleon transfer reactions $(\vec{d},\alpha)$ 
not only offer a powerful tool to help establish the spin and parity 
assignments of the energy levels in the odd-odd nucleus $^{196}$Au 
\cite{wirth}, but also provide a sensitive test of neutron-proton 
correlations in the wave functions. The symmetry structure of the 
model gives rise to selection rules and parameter independent 
predictions of ratios of spectroscopic strenghts which only depend 
on the nuclear structure part, and not on factors that arise from 
the kinematical part. A comparison with experimental data shows 
a surprisingly good overall agreement with the predictions of the 
supersymmetry scheme and, in this way, lends further support to the 
validity of the supersymmetry scheme in atomic nuclei, especially 
in the Pt-Au mass region. More tests involving the $^{194}$Pt and 
$^{196}$Au nuclei \cite{BBF,JPA} as well as nuclei in other parts of 
the nuclear mass table \cite{algora} are currently underway.

\begin{acknowledgments}
This work was supported in part by Conacyt. We are grateful to 
G. Graw for sharing the new experimental data on the 
$^{198}\textrm{Hg}(\vec{d},\alpha){}^{196}\textrm{Au}$ pick-up
reaction prior to publication. Enlightening discussions with 
J. G\'omez-Camacho and P. Van Isacker are gratefully acknowledged.
\end{acknowledgments}


\begin{thebibliography}{99}

\bibitem{IBM} 
A. Arima and F. Iachello, 
Phys. Rev. Lett. {\bf 35}, 1069 (1975); {\bf 40}, 385 (1978).

\bibitem{IBFM} 
F. Iachello and O. Scholten, 
Phys. Rev. Lett. {\bf 43}, 679 (1979).
 
\bibitem{FI} 
F. Iachello, 
Phys. Rev. Lett. {\bf 44}, 772 (1980). 

\bibitem{quartet}
P. van Isacker, J. Jolie, K. Heyde and A. Frank, 
Phys. Rev. Lett. {\bf 54}, 653 (1985).

\bibitem{metz}
A. Metz, J. Jolie, G. Graw, R. Hertenberger, J. Gr\"oger, 
C. G\"unther, N. Warr and Y. Eisermann, 
Phys. Rev. Lett. {\bf 83}, 1542 (1999).

\bibitem{Glendenning}
N.K. Glenndenning, 
{\it Direct Nuclear Reactions}, 
(Academic Press, New York, 1983).

\bibitem{wirth}
H.-F. Wirth, G. Graw, S. Christen, Y. Eisermann, A. Gollwitzer, 
R. Hertenberger, J. Jolie, A. Metz, O. M\"oller, D. Tonev and B.D. Valnion, 
Phys. Rev. C {\bf 70}, 014610 (2004).

\bibitem{baha}
A.B. Balantekin, I. Bars, R. Bijker and F. Iachello, 
Phys. Rev. C {\bf 27}, 1761 (1983);\\
H.Z. Sun, A. Frank and P. Van Isacker, 
Phys. Rev. C {\bf 27}, 2430 (1983). 

\bibitem{IK}
F. Iachello and S. Kuyucak, 
Ann. Phys. (N.Y.) {\bf 136}, 19 (1981).

\bibitem{IFS}
P. Van Isacker, A. Frank and H.-Z. Sun, 
Ann. Phys. (N.Y.) {\bf 157}, 183 (1984).

\bibitem{BI}
R. Bijker and F. Iachello, 
Ann. Phys. (N.Y.) {\bf 161}, 360 (1985).

\bibitem{PvI}
P. Van Isacker, 
J. Math. Phys. {\bf 28}, 957 (1987).

\bibitem{BBF}
J. Barea, R. Bijker and A. Frank, in preparation. 

\bibitem{au196}
J. Gr\"oger, J. Jolie, R. Kr\"ucken, C.W. Beausang, M. Caprio, R.F. Casten, 
J. Cederkall, J.R. Cooper, F. Corminboeuf, L. Genilloud, G. Graw, 
C. G\"unther, M. de Huu, A.I. Levon, A. Metz, J.R. Novak, N. Warr and 
T. Wendel, 
Phys. Rev. C {\bf 62}, 064304 (2000).

\bibitem{JPA}
J. Barea, R. Bijker and A. Frank, 
J. Phys. A: Math. Gen. {\bf 37}, 10251 (2004).

\bibitem{algora} 
A. Algora, J. Jolie, Zs. Dombradi, D. Sohler, Zs. Podoly\'ak and 
T. F\'enyes, 
Phys. Rev. C {\bf 67}, 044303 (2003).

\end{thebibliography}
\end{document}